# A high-performance nitrogen-rich ZIF-8-derived Fe-Co-NC electrocatalyst for the oxygen reduction reaction

Yuqin Wang[a,b], Lizi He[a,b*], Ning Han[a,b]

a. Key Laboratory of Electromagnetic Processing of Materials, Ministry of Education, Northeastern University, Shenyang 110819, PR China

b. School of Materials Science and Engineering, Northeastern University, Shenyang, 110819, China

*Corresponding author, Email address: helizi@epm.neu.edu.cn

## Abstract

It is of great practical significance to explore and prepare cheap, high-performance and stable catalysts for oxygen reduction reaction (ORR), but it is still in progress at present. Herein, a straightforward evaporation-pyrolysis strategy is designed for the preparation of 3D porous electrocatalys (denoted as Fe-Co-NC, which represented carbonization products at 900°C) made of carbon nanoparticles combined with metallic FeCo doped N enricheds bridged carbon nanotube by carbonization of a pristine ZIF-8 as highly efficient and durable ORR electrocatalysts. The obtained Fe-Co-NC structure possess 3D open porous texture, abundant active sites, desired nitrogen bonding type and high specific surface area, providing them with excellent ORR activity. The optimal-performing Fe-Co-NC catalyst presents an outstanding ORR performance in terms of a high onset potential ($E_{onset}$) of 0.96 V (vs. RHE) and half-wave potential ($E_{1/2}$) of 0.86 V (vs. RHE), respectively, possible to rival those of the commercially available Pt/C in an alkaline electrolyte solution. Besides, the Fe-Co-NC catalyst manifests better stability than those of commercially available Pt/C, which is of importance for the contemplation and characterization of novel electrocatalysts originated from non-precious metals.

## 1. Introduction

With the aggravation of global energy crisis and environmental degradation, it is particularly important to develop new energy conversion systems. Oxygen reduction reaction (ORR) is a key process in fuel cells and metal air cells. However, the slow reaction speed and high reaction energy barrier seriously hinder its practical application. Platinum and its alloys have become the standard because of their excellent ORR catalytic performance, but the high cost, small reserves and poor stability have posed a major obstacle to large-scale commercialization.[1] Therefore, it is extremely urgent to explore efficient and low-cost electrocatalysts with competitive advantages such as excellent ORR performance and high stability to replace pt-based catalysts.[2]

Among various non-precious metal catalysts, carbon-based composite materials have emerged as key research focus due to their superior conductivity, large specific surface area, stable structure and tunability.[3,4] These materials have enhanced ORR activity by doping heteroatoms (such as N, P, S)[5-7] and transition metals (such as Fe, Co, Ni)[8-10] to modify their electronic and geometric structures. Studies have shown that transition metal and nitrogen-doped carbon materials, particularly those with M-Nx active centers (such as Fe-$N_x$, Co-$N_x$), are critical to ORR properties.[11] However, traditional carbon-based catalysts face serious challenges in the process of high temperature pyrolysis, in which weak chemical bonds between metal atoms and carbon matrix usually lead to metal atoms aggregating into metal clusters, oxides or carbides with low activity, thus significantly reducing the activity and stability of catalysts.[12] Therefore, to improve the catalyst performance, it is necessary to effectively capture and anchor metal atoms during high temperature calcination to prevent aggregation. Metal-organic frameworks (MOFs), especially ZIF-8, have been widely used to synthesize M-N-C catalysts in recent years due to their stable structure, high porosity, and large surface area.[8] ZIF-8 is a three-dimensional porous skeleton formed by the coordination of zinc ions and nitrogen in dimethylimidazole, which is easy to synthesize and has excellent thermal and chemical stability, and is a commonly used catalyst precursor. Studies have shown that ZIF-derived mono-metallic carbon-based catalysts, such as Fe-N-C, Co-N-C, and Cu-N-C, exhibit excellent ORR performance.[11] However, their stability and selectivity remain suboptimal. For example, in the alkaline ORR process, Fenton reaction easily occurs between $H_2O_2$ by-products and leached Fe ions in Fe-N-C catalyst, which leads to the generation of reactive oxygen species (ROS), which erodes the support and aggravates carbon corrosion.[13,14] One strategy for stabilizing the Fe-N-C catalysts is to minimize the concentration of ROS from Fenton reactions by eliminating the ROS or reducing the reactant $H_2O_2$.[15] Recent studies have explored the synergistic effect of introducing a second transition metal with iron to enhance the stability and ORR activity of the catalyst.[16] The combination of iron and other metal centers creates diverse adsorption sites, potentially breaking the linear relationship between the adsorption energies of different oxygen species and enabling more complex catalytic reactions.[17] For example, Zhang et al.[18] recently reported a distinct two-electron pathway to reduce unwanted $H_2O_2$ to $H_2O$ through Fe-Cu diatomic sites. Density functional theory (DFT) calculations indicate that Fe-Cu dual-atom catalysts (DAC) have a higher energy barrier for Fenton-like reactions and a lower energy barrier for hydrogen peroxide reduction reactions (HPRR) due to the synergistic effects of the Fe-Cu metal pairs. Similarly, Niu et al.[15] demonstrated that Pt-Fe-NC catalyst exhibited excellent ORR activity, with half-wave potential of 0.93 V vs.RHE and $H_2O_2$ yield of only 0.6% at 0.6 V. The catalyst also showed remarkable durability, with a potential decay of only 7 mV after 50,000 potential cycles.

Although the ORR activity of Fe-NC catalysts has made remarkable progress, the research on improving its stability and ORR performance by doping different transition metals is still relatively limited.[17] Building on previous work, this study synthesized Fe-Cu-ZIF8, Fe-Ni-ZIF8, and Fe-Co-ZIF8 composite precursors by mixing zinc salts with other transition metal salts in methanol at room temperature. During the pyrolysis at 900°C, zinc ions selectively evaporated, with some being replaced by other transition metal ions, increasing the spacing between metal atoms and inhibiting the formation of metal clusters. Electrochemical testing revealed that the half-wave and onset potentials of Fe-Co-NC, Fe-Ni-NC, and Fe-Cu-NC were all higher than those of Fe-NC. Fe-Co-NC exhibited the highest half-wave potential of 0.861 V (vs. RHE) and maintained 89.3% of its initial performance after 15,000 seconds in chronoamperometric test, compared with only 87.9% and 73.9% for commercial 20% Pt/C and Fe-NC, respectively.In conclusion, compared with single metal doped catalysts, the bimetallic doped catalysts prepared in this study show significant improvement in long-term stability and ORR activity, highlighting their potential for practical applications.

## 2. Result and discussion

### 2.1. *Materials*

Anhydrous methanol (superior purity, Sinopharm), $Zn(NO_3)_2 \cdot 6H_2O$, $Cu(NO_3)_2 \cdot 3H_2O$, $Ni(NO_3)_2 \cdot 6H_2O$, $Fe(NO_3)_3 \cdot 9H_2O$, $Co(NO_3)_2 \cdot 6H_2O$ were all from Sinopharm, 2-methylimidazole (Aladdin), Nafion solution (DuPont), commercial Pt/C catalyst ( Tanaka Precious Metals Group), potassium hydroxide (Sinopharm), deionized water.

### 2.2. *Catalyst preparation*

Synthesis of Fe-ZIF8: First, 2.97 g of $Zn(NO_3)_2 \cdot 6H_2O$ and 0.2 g of $Fe(NO_3)_3 \cdot 9H_2O$ were weighed and dissolved in 50 mL of anhydrous methanol, labelled as solution A. Next, 5 g of 2-methylimidazole were weighed and dissolved in 100 mL of anhydrous methanol, labelled as solution B. To ensure the uniform dispersion of the solution, solutions A and B were sonicated separately for 5 min. Then, solution B was poured into solution A and stirred at room temperature for 24 h to promote the reaction. Upon completion of the reaction, the solid product was separated by centrifugation, washed three times with methanol solution, then filtered, and the solid obtained was placed in a vacuum drying oven and dried at 60 °C overnight to obtain the Fe-ZIF8 precursor.

Synthesis of Fe-Co-ZIF8, Fe-Ni-ZIF8 and Fe-Cu-ZIF8: On the basis of the synthesis of Fe-ZIF8, 0.15 g of $Co(NO_3)_2 \cdot 6H_2O$, 0.09 g of $Ni(NO_3)_2 \cdot 5H_2O$, and 0.12 g of $Cu(NO_3)_2 \cdot 3H_2O$ were added to solution A, respectively, to prepare Fe-Co -ZIF8, Fe-Ni-ZIF8 and Fe-Cu-ZIF8. These solutions were processed by the same steps of stirring, centrifugationing, washing and drying to finally obtain the corresponding precursor materials.

Synthesis of Fe-NC, Fe-Cu-NC, Fe-Ni-NC and Fe-Co-NC: The dry precursor powders were fully milled separately, transferred to a corundum porcelain boat with lid and placed in the tube furnace. Under the protection of nitrogen ($N_2$) inert gas, the furnace was heated to 900 °C at a heating rate of 5 °C/min and kept for a holding time of 2 h. Upon completion, the samples were naturally cooled to room temperature, and the black powders obtained were named Fe-NC, Fe-Cu-NC, Fe-Ni-NC, and Fe-Co-NC, respectively, and were sufficiently ground to ensure the homogeneity of the catalysts.

Synthesis of NC: Undoped metal ZIF8 was synthesized without adding any metal salt for comparative experiment.

### 2.3 *Characterization*

Structural analysis of the catalysts was carried out by X-ray diffractometer (XRD, Rigaku Smart Lab) with a scanning range of 10-80° and a scanning speed of 10°$min^{-1}$. The chemical element composition and valence state of the material surface were analyzed using X-ray photoelectron spectroscopy (XPS, Thermo Escalab 250Xi). The real surface areas of the catalysts were measured by BET method based on nitrogen adsorption on an ASAP 2460 analyzer (BET, McMurray Tic Instruments Ltd). Raman spectrometer model DXR2 from Thermo Fisher was used to measure the concentration of carbon defects and graphitization information of the catalysts. SEM images of the catalysts were recorded using a scanning electron microscope (SEM, Zeiss Sigma 300) operated at 15 KV.

All electrochemical tests were performed on a CHI 760e electrochemical workstation based on a three-electrode system, with the test electrolyte being a 0.1 mol/L KOH solution. Ag/AgCl electrode was used as the reference electrode, a metallic platinum sheet was used as the counter electrode, and a rotating disk electrode (RDE) coated by a droplet of catalyst ink was used as the working electrode. Specifically, 8 mg of catalyst powder was dispersed in 950 μL of ethanol and 50 μL of deionized water and sonicated for 1 h to ensure uniform dispersion. Subsequently, 9 μL of the catalyst ink was aspirated using a pipette gun, and uniformly dripped onto a glassy carbon electrode with a diameter of 5 mm, which was allowed to dry naturally and then prepared for use. The catalyst loading on each working electrode was 0.36 mg/cm², while the loading of Pt/C catalyst was 0.18 mg/cm². Measurements were carried out using linear scanning voltammetry (LSV) at a scanning speed of 5 mV/s and 1600 rpm with scanning voltages ranging from 0.17 V to 1.07 V (vs. RHE), with oxygen being introduced for 30 min prior to the test and a continuous supply of airflow during the test. To further calculate the number of electrons transferred for the catalytic reaction, LSV tests at different speeds (400 rpm, 900 rpm, 1600 rpm, 2500 rpm) were performed. Based on the Koutecky-Levich (K-L) equation, the linear relationship between current density and rotational speed at different potentials was calculated to derive the electron transfer number. In addition, the chronoamperometry (CA) method was used to evaluate the long-term stability of the catalyst, which was tested for 15,000 s at an initial potential of 0.7 V (vs. RHE). The rotating ring-disk electrode (RRDE) was used to measure the electron transfer number and H2O2 yield experiments were performed at a scanning rate of 5 mV/s, the ring voltage was set to 1.3 V (vs. RHE), and the calculation of the $H_2O_2$ yield and the electron transfer number (n) were determined according to the following two equations:

$$H_2O_2\% = \frac{200 I_R/N}{I_R + I_D} \qquad (1)$$

$$n = \frac{4 I_D}{I_D + I_R/N} \qquad (2)$$

Where $I_R$ and $I_D$ are the ring and disc currents, respectively, and $N = 0.37$ is the collection efficiency of the circular ring.

## 3. Results and Discussion

The electrochemical ORR behaviour of NC, Fe-NC, Fe-Cu-NC, Fe-Co-NC, Fe-Ni-NC and Fe-NC were first evaluated in 0.1 M $O_2$-saturated KOH solution by linear sweep voltammetry (LSV) with the identical mass loading (Fig. 1a). The half-wave potentials of NC, Fe-Cu-NC, Fe-Co-NC, Fe-Ni-NC and Fe-NC were calculated to be 0.780, 0.848, 0.861, 0.843 and 0.819 V, respectively (Fig. 1b). Obviously, Fe-Co-NC presented an enhanced ORR ability than other catalysts, which can rival with the benchmark Pt/C (0.853 V) (Fig. 1c). Moreover, Fe-Co-NC also shows the highest onset potential of 0.96 V than that of other catalysts. In order to further validate the mechanism of the ORR process, we have plotted the Tafel diagrams derived from LSV curves of all catalysts in Fig. 1d. Fe-Co-NC delivers a Tafel slope of 67 mV·dec$^{-1}$, which is much smaller than the Tafel slope (67 mV·dec$^{-1}$) of Pt/C, suggesting that the first electron transfer is the dominating the rate-determining step of the ORR on the Fe-Co-NC.

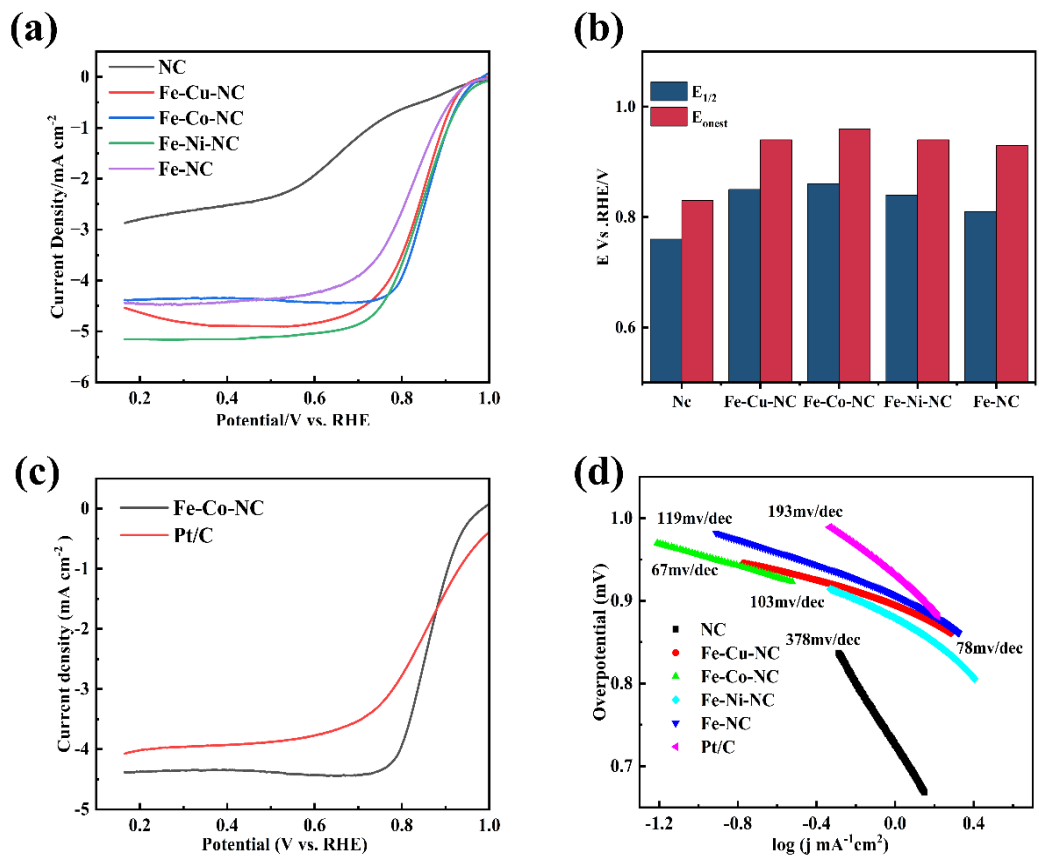

Fig. 1. (a) LSV curves of NC、Fe-NC、Fe-Cu-NC, Fe-Co-NC, Fe-Ni-NC, Fe-NC in $O_2$-saturated 0.1 mol/L KOH solution at 1600 rpm. (b) half-wave potentials and onset potentials of NC, Fe-NC, Fe-Cu-NC, Fe-Co-NC, Fe-Ni-NC, Fe-NC statistical plots. (c) Tafel slope plots for NC, Fe-NC, Fe-Cu-NC, Fe-Co-NC, Fe-Ni-NC, Fe-NC. (d) LSV curves for Fe-Co-NC and Pt/C at 1600 rpm.

The electrocatalytic kinetics of Fe-Co-NC was further investigated under variable rotation rates ranging from 400 to 2500 rpm. As shown in fig. 2a, with the increase of rotating speed, the concentration polarization effect decreases, which leads to the increase of limiting current density and the acceleration of diffusion rate of reactants on the catalyst surface, indicating that the Fe-Co-NC carry porous channels. Koutecky-Levich (K-L) pro-files calculated from the RDE LSV curves were displayed in Fig. 2b. The corresponding K-L profiles have a benign linear relationship and persistent slope, hinting first-order reaction kinetics toward the concentration of dissolved oxygen. In addition, the fitted profiles indicated that the average number of transferred electrons n value was calculated to be approximately 3.98 on account of K-L formula, revealing that the Fe-Co-NC catalyst ideally executes ORR through the 4e$^-$ pathway.

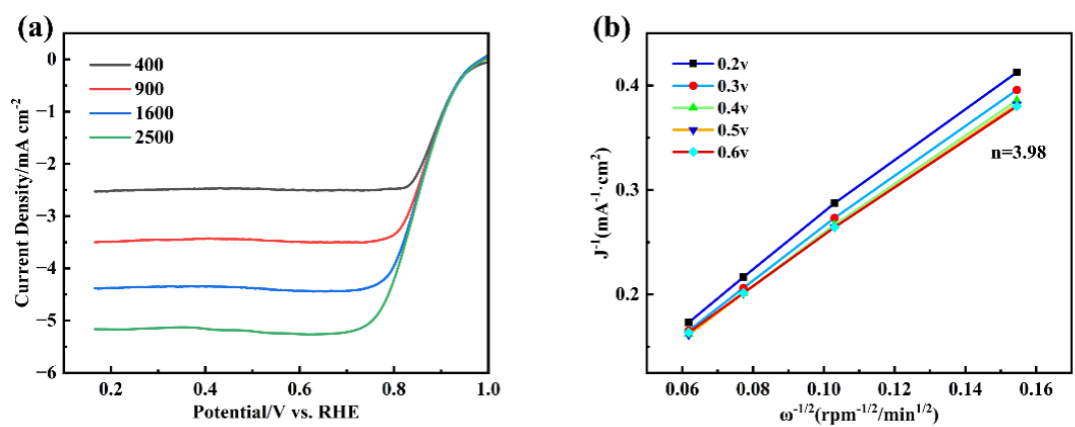

Fig. 2. (a) LSV plots of Fe-Co-NC at 400 rpm to 2500 rpm in 0.1 M KOH saturated with $O_2$. (b) K-L plots at 0.2v to 0.6v potentials.

To further confirm the above result, the rotating ring disk electrode (RRDE) trials were operated to detect the evolution of the reaction intermediate ($H_2O_2$) and to ascertain the electron transfer number during the ORR. Fig. 3a displays that Fe-Co-NC affords minor ring current density and major disk current density, which means that the reduction of oxygen is more likely to occur on Fe-Co-NC catalyst than the oxidation of peroxide, attesting a superior selectivity towards ORR. The number of transferred electrons and yield of $H_2O_2$ was calculated according to the formulas (1) and (2). Fig. 3b clearly shows that the n value of the Fe-Co-NC catalyst was located between 3.65 and 3.75 and the $H_2O_2$ yield observed on the sample was less than 3% in a potential range (0.2–0.8 V), which was compatible with the results obtained from K-L plots. The above results confirm the ideal $4e^-$ path to $H_2O$ was dominating in the ORR process of Fe-Co-NC.

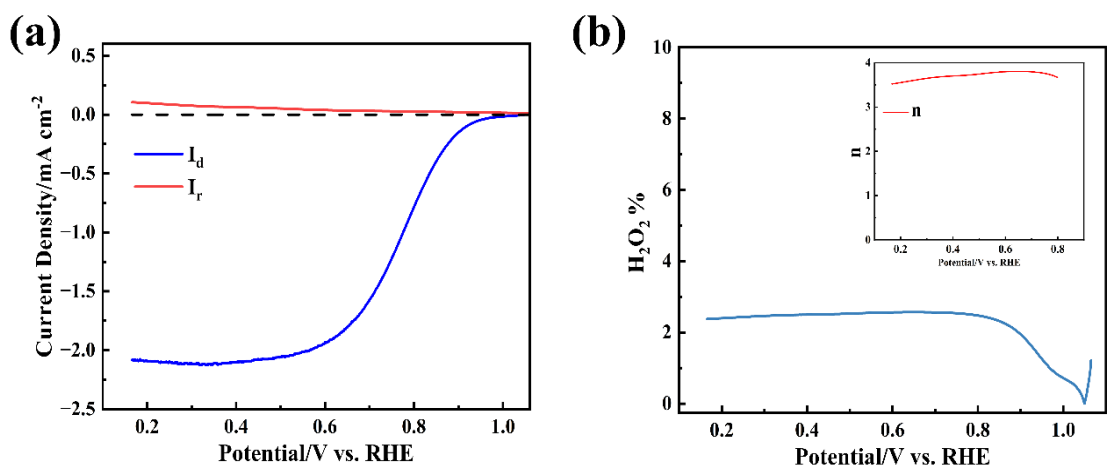

Fig.3. (a) Plot of $I_r$ and $I_d$ of Fe-Co-NC. (b) Plot of electron transfer number and $H_2O_2$ yield of Fe-Co-NC.

Apart from the catalytic activity, the stability of an ORR catalyst are important concerns for its applications in the actual market. In an alkaline environment, the Fe-Co-NC catalyst reveals excellent durability in contrast to that of the commercially available Pt/C catalyst. The cycling measurement was performed by a chronoamperometric trial for 15000 s in 0.1 M KOH solution. The chronoamperometry profile for Fe-Co-NC presents a mild variation with 89.3% retention after operating for 15000s, whereas the chronoamperometry profile of the

Fe-NC and commercially available Pt/C declined remarkably, them only preserves 73.90% and 87.90% of initial current density after the same cycling tests (Fig. 4).

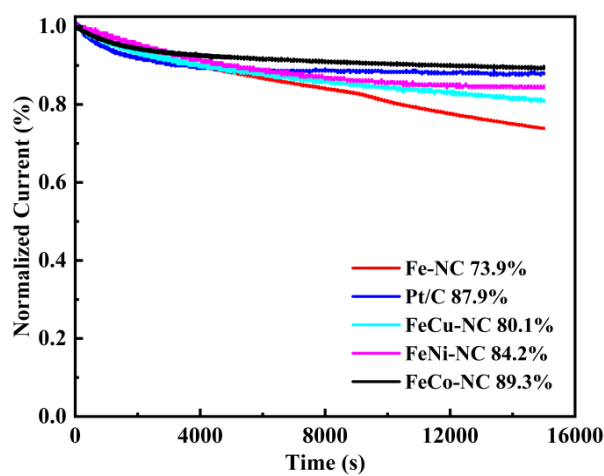

Fig. 4. Timing current response plots for Fe-NC, Fe-Cu-NC, Fe-Co-NC, Fe-Ni-NC, Fe-NC and Pt/C.

In order to study the morphology of the catalysts, scanning electron microscopy (SEM) was used to characterize the Fe-Co-NC, Fe-Cu-NC and Fe-Ni-NC materials. As shown in Fig. 5a, the average diameter of Fe-Co-NC nanoparticles is between 500 and 600 nm, showing a regular dodecahedral structure. By further observing Fig. 5b, we can see that small size nanoparticles with a diameter of about 200 nm are distributed around the larger size Fe-Co-NC nanoparticles. In addition, a large number of spiral carbon nanotubes (SWCNTs) were grown on the surface of the material. The formation of carbon nanotubes is usually due to the transition metal nanoparticles (such as Fe, Co, Ni, Cu, etc.) on the surface of materials catalyzing the formation of carbon caps around carbon atoms, then inducing the growth of carbon nanotubes. This phenomenon indicates that the activity of metal elements in materials promotes the formation of carbon nanotube structure. It is worth noting that SWCNT plays an important role in promoting electron transfer and further improves the catalytic efficiency of ORR reaction. This discovery provides a strong structural basis for understanding the high efficiency of Fe-Co-NC catalyst, and points out its application potential in carbon-based electrocatalytic system.[19] Figs. 6c and 6d show the morphology of the Fe-Cu-NC catalyst, which also has a definite dodecahedral shape, but has a slightly different surface structure compared to Fe-Co-NC. Copper doping seems to lead to a decrease in the growth of carbon nanotubes, although some still exist. This indicates that although Cu contributes to graphitization and provides active sites for ORR, it may not be as effective as Co in catalyzing the growth of SWCNTs. However, dodecahedral structure is still very beneficial to ORR because it provides large surface area and multiple active sites. Compared with Fe-Co-NC and Fe-Cu-NC, the Fe-Ni-NC catalysts shown a more irregular surface morphology (Fig. 6e and 6f). Nanoparticles exhibit a coarser, more heterogeneous structure, although they still maintain a certain degree of dodecahedral geometry. It is worth noting that the surface of Fe-Ni-NC catalyst seems to contain less single-walled carbon nanotubes, which may be due to the different catalytic behaviors of Ni in promoting the growth of carbon nanotubes.

Nevertheless, rough and irregular surfaces can provide a high density of active sites, potentially compensating for the reduction of SWCNTs in enhancing ORR performance. In summary, each catalyst presents unique morphological characteristics that contribute to their ORR activity. The Fe-Co-NC catalyst, with its abundant SWCNTs and well-defined structure, offers excellent electron transfer capabilities, while the Fe-Cu-NC catalyst benefits from a similar structure with slightly fewer SWCNTs. In contrast, the Fe-Ni-NC catalyst, despite its more irregular morphology, still provides a high density of active sites due to its rougher surface, which may also contribute to ORR efficiency. The comparison of these three catalysts highlights the role of metal dopants in influencing both the structural and electrochemical properties of the materials.

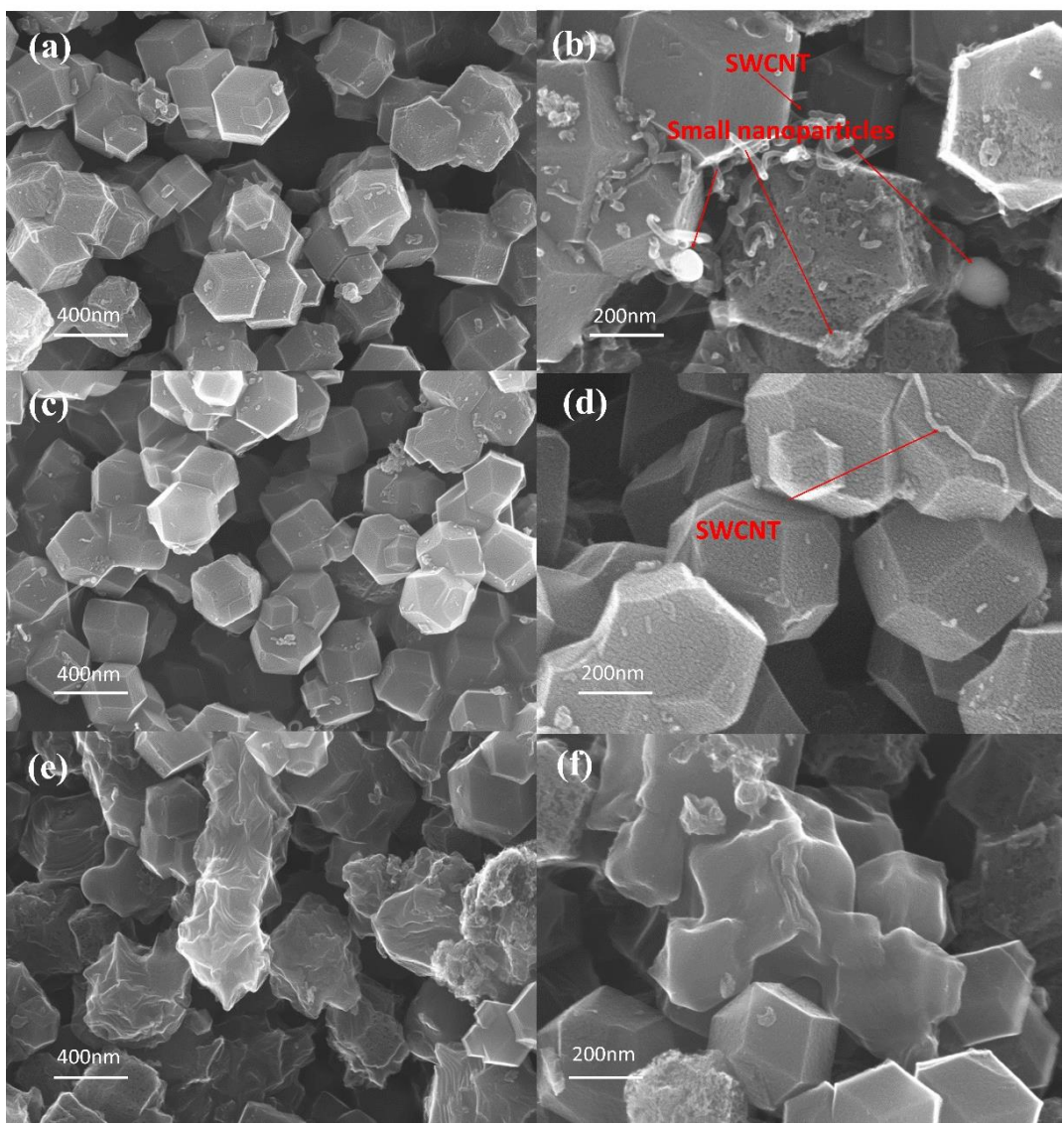

Fig. 5. (a) (b) Scanned images of Fe-Co-NC. (c) (d) Scanned images of Fe-Co-NC. (e) (f) Scanned images of Fe-Co-NC.

Fig. 6a presents the Raman spectrum of the Fe-Co-NC catalyst, where the D and G bands are observed at approximately 1340 cm$^{-1}$ and 1590 cm$^{-1}$, respectively.[20] These peaks are associated with defect sites in amorphous carbon (D band) and graphitized carbon (G band). Specifically, the D band signifies the presence of defects or amorphous carbon content, while the G band reflects the degree of graphitization in the carbon material. The intensity ratio of ID/IG serves as a crucial indicator of the graphitization level. A lower ID/IG ratio implies a higher proportion of graphitized carbon and enhanced conductivity due to reduced defect concentration.[20] The calculated $I_D/I_G$ ratio for the Fe-Co-NC catalyst, as shown in Fig. 6a, is 1.04, which is slightly lower than the ratio of 1.09 observed in NC material without metal doping. This suggests that the incorporation of Fe and Co enhances the material's graphitization, consequently improving its electrical conductivity. This increase in electrical conductivity is essential to enhanced the electrocatalytic activity of the catalyst in the oxygen reduction reaction (ORR), indicating that the doped metals effectively modifies the catalyst.

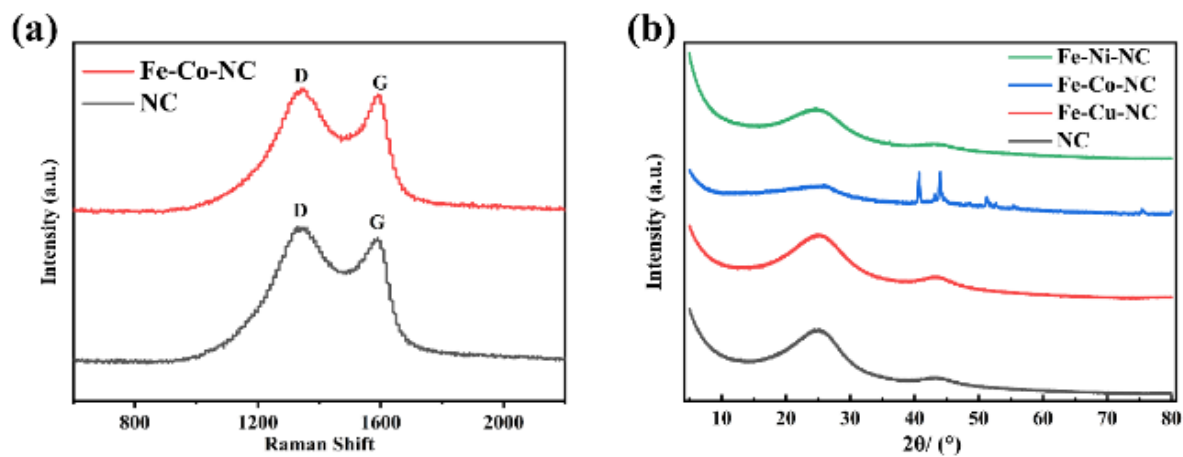

Fig. 6. (a) Raman image of Fe-Co-NC, NC. (b) XRD images of Fe- Ni -NC, Fe-Co-NC, Fe-Cu-NC and NC.

The X-ray diffraction (XRD) patterns of Fe-Co-NC, Fe-Ni-NC, and Fe-Cu-NC are shown in Fig. 6b. The diffraction peaks at approximately 2θ values of 25° and 42.4° correspond to the (002) and (100) crystal planes of graphitic carbon, respectively,[21] indicating that all these catalysts have a certain degree of graphitic carbon structure. Additionally, three peaks attributed to the (111), (200), and (220) crystal planes of metallic cobalt (Co) appear at 2θ values of 44.3°, 51.5°, and 75.8°, respectively.[22] In the XRD patterns of Fe-Co-NC, confirming cobalt ions are reduced to metal forms during the pyrolysis of Fe-Co-NC.

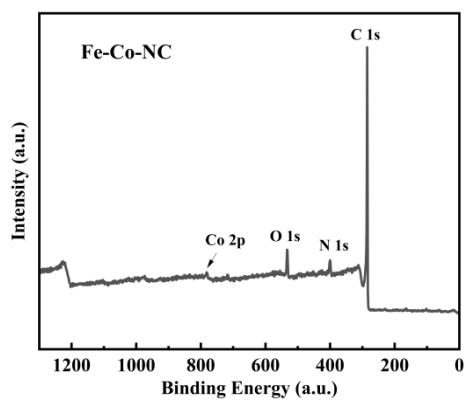

Fig .7. XPS mapping of Fe-Co-NC

To further investigate the elemental composition and chemical state of the Fe-Co-NC catalyst, X-ray photoelectron spectroscopy (XPS) was utilized for characterization. Fig. 7 presents the complete XPS spectrum of Fe-Co-NC, revealing clear signals from the elements carbon (C), nitrogen (N), oxygen (O), and cobalt (Co). To analyze the chemical state of each element in detail, the fine spectra of C 1s and N 1s were fitted to identify the splitting peaks. Fig. 8a illustrates the split-peak analysis for C 1s, where binding energy peaks at 284.8 eV, 286.4 eV and 289.1 eV correspond to C-C, C=N and O=C-C bonds, respectively.[23] The existence of these carbon bonds suggests various structural forms of carbon within the materials. Notably, the presence of nitrogen in the C=N bond modifies the charge density of adjacent carbon atoms, thereby increasing the number of active sites[22]. The pronounced C-C peak indicates a significant degree of graphitization in the catalyst, which is crucial for enhancing its electrical conductivity.

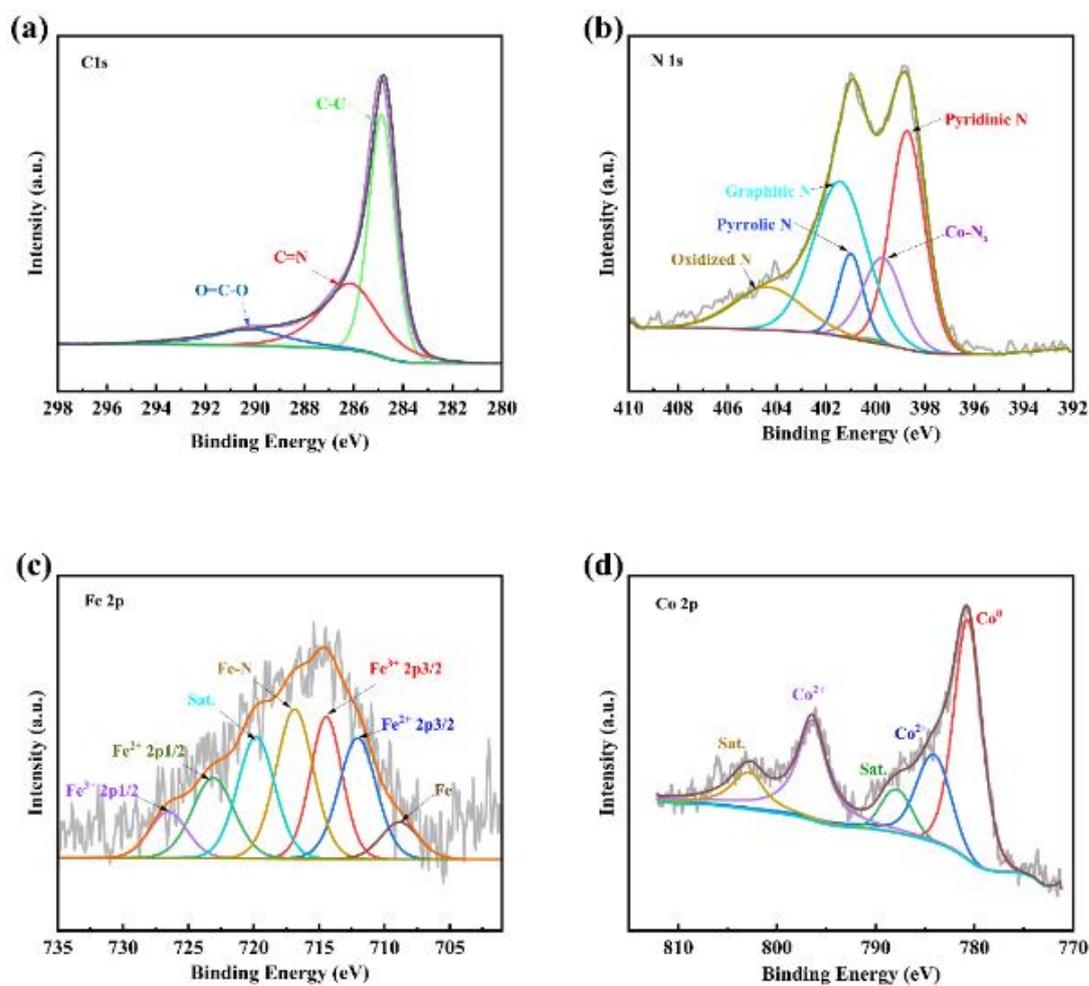

Fig. 8. High-resolution spectra of Fe-Co-NC catalysts: (a) C 1s; (b) N 1s; (c) Co 2p; (d) Fe 2p.

Fig. 8b displays the peak splitting results for N 1s, indicating that nitrogen exists in five different chemical states in Fe-Co-NC: oxidized nitrogen (404.2 eV), graphitic nitrogen (401.3 eV), pyrrolic nitrogen (400.9 eV), metallic nitrogen (399.61 eV) and pyridine nitrogen (392.8 eV).[23] Among them, graphitic nitrogen, metallic nitrogen and pyridine nitrogen serve as primary sites for the adsorption and activation of oxygen molecules, playing a vital role in the oxygen reduction reaction (ORR).[22] Pyridine nitrogen enhances the adsorption of oxygen on the catalyst surface through synergistic interactions, thus improving the electrocatalytic activity of the catalyst. Conversely, the presence of metallic nitrogen boosts ORR activity by modulating the electron density and its interaction with oxygen molecules.[24] The distribution of these nitrogen species highlights their functional diversity and essential contribution to the ORR performance of Fe-Co-NC catalysts. Furthermore, the peak position of Co-$N_x$ corroborates the successful doping of cobalt. Fig. 8d shows the high-resolution spectrum of Co 2p, with peaks at 781.8 eV and 795.9 eV corresponding to the Co $2p_{3/2}$ and Co $2p_{1/2}$ orbitals of cobalt ions ($Co^{2+}$), alongside satellite peaks at 785.2 eV and 801.9 eV. Importantly, a peak at 779.2 eV corresponding to cobalt metal ($Co^0$) was also detected.[25] These findings demonstrates that cobalt in Fe-Co-NC exists in various oxidation states and chemical forms, indicating the presence of Co-O and Co-N bonds, which contribute to the diverse active centers for enhancing the electrocatalytic performance of the catalyst.[21]

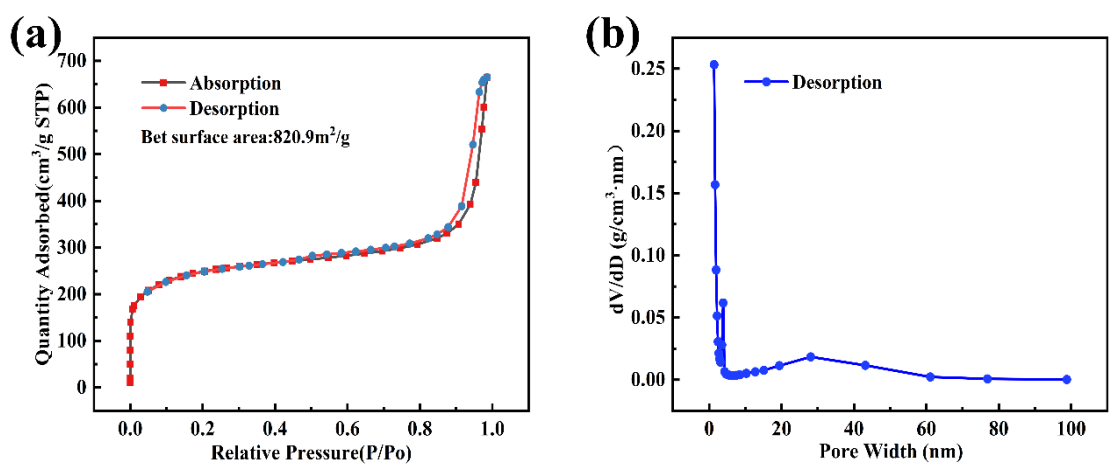

Fig. 9. (a) Nitrogen adsorption and desorption curves of Fe-Co-NC. (b) Pore size distribution of Fe-Co-NC

Table 1. Structural properties of Fe-Co-NC catalysts

| Sample | BET Surface Area ($m^2/g$) | Volume of pores ($cm^3/g$) | Desorption average pore diameter (nm) |
|---|---|---|---|
| Fe-Co-NC | 820.9 | 1.02 | 4.08 |

Fig. 9a presents the nitrogen adsorption and desorption isotherms of Fe-Co-NC, exhibiting a characteristic hysteretic type IV isotherm, indicating mesoporous structure. This mesoporous framework facilitates substantial material transport channels during the oxygen reduction reaction (ORR), thereby enhancing the diffusion of reactants and products and ultimately improving the efficiency of the ORR. To elucidate the pore structure, Fig. 9b illustrates the pore size distribution of Fe-Co-NC. The data reveal that the pore sizes primarily range from 2 to 50 nm, confirming the presence of both micropores and mesopores. Such porous architecture is advantageous for increasing the specific surface area and exposing a greater number of active sites, which in turn provides more reaction sites for the ORR and significantly boosts the electrocatalytic activity of the catalyst.[23] The structural characteristics of the Fe-Co-NC material are summarized in Table 1, which shows a high specific surface area of 820.9 $m^2·g^{-1}$, a pore volume of 1.02 $cm^3·g^{-1}$, and an average desorption pore size of 4.08 nm. This high specific surface area and proper pore size distribution enables Fe-Co-NC to effectively enhance the rapid transport of reactive species during the ORR while providing numerous active sites for the reaction.[26] These structural attributes underscore the potential applicability of Fe-Co-NC catalysts in the field of electrocatalysis.[27]

# 4 Conclusion

In this study, bimetallic doped non-precious metal catalysts were developed, which exhibit excellent oxygen reduction activity and longtime stability, especially the Fe-Co-NC catalyst with good performance. Its half-wave potential is 0.861 V, which keeps 89.3% of the initial performance in chronoamperometric method. Meanwhile, the physical characterization results shows that the elements of Fe and Co were successfully doped, with high graphitization degree, good electrical conductivity and rich pore structure, which results in the high ORR catalytic performance of Fe-Co-NC. The synergistic effect of the bimetal significantly enhanced the electrocatalytic ability of catalysts, which provides a valuable reference for the design of high-efficiency and low-cost non-precious metal catalysts in the future.

Declaration of Competing Interest
The authors declare that they have no conflict of interest.
Acknowledgments
This work was supported by the National Key R&D Program of China (Grant No. 2023YFB3710401). Special thanks are due to the instrumental or data analysis from Analytical and Testing Center, Northeastern University.

**References**
[1] J. Kong, Y. H. Qin and T. L. Wang. Photodeposition of Pt nanoparticles onto TiO2@CNT as high-performance electrocatalyst for oxygen reduction reaction. *International Journal of Hydrogen Energy.* **3**, 45 (2020), pp. 1991-1997.
[2] Z. Li, Q. Gao and W. Qian. Ultrahigh Oxygen Reduction Reaction Electrocatalytic Activity and Stability over Hierarchical Nanoporous N-doped Carbon. *Scientific Reports.* **1**, 8 (2018), pp. 2863.
[3] M. Liu, L. Wang and K. Zhao. Atomically dispersed metal catalysts for the oxygen reduction reaction: synthesis, characterization, reaction mechanisms and electrochemical energy applications. *Energy & Environmental Science.* **10**, 12 (2019), pp. 2890-2923.
[4] He, S. Liu and C. Priest. Atomically dispersed metal–nitrogen–carbon catalysts for fuel cells: advances in catalyst design, electrode performance, and durability improvement. *Chemical Society Reviews.* **11**, 49 (2020), pp.: 3484-3524.
[5] H. Sun, M. Wang and S. Zhang. Boosting Oxygen Dissociation over Bimetal Sites to Facilitate Oxygen Reduction Activity of Zinc-Air Battery. *Advanced Functional Materials.* **4**, 31 (2021).
[6] Y. Peng, B. Lu and S. Chen. Carbon-Supported Single Atom Catalysts for Electrochemical Energy Conversion and Storage. *Advanced Materials.* (2018).
[7] C. H. Choi, M. Kim and H. C. Kwon. Tuning selectivity of electrochemical reactions by atomically dispersed platinum catalyst. *Nature Communications.* **1**, 7 (2016).
[8] H. Zhang and S. H. Wang. Single Atomic Iron Catalysts for Oxygen Reduction in Acidic Media: Particle Size Control and Thermal Activation. *Journal of the American Chemical Society.***40**, 13

9 (2017), pp. 14143-14149.

[9] J. Wang, Z. Huang and W. Liu. Design of N-Coordinated Dual-Metal Sites: A Stable and Active Pt-Free Catalyst for Acidic Oxygen Reduction Reaction. *Journal of the American Chemical Society*. **48**, 139 (2017), pp. 17281-17284.

[10] D. Yu, Y. Ma and F.Hu. Dual-Sites Coordination Engineering of Single Atom Catalysts for Flexible Metal–Air Batteries. *Advanced Energy Materials*. **30,** 11 (2021).

[11] C. Zhao, B. Li and J. Liu. Intrinsic Electrocatalytic Activity Regulation of M-N-C Single-Atom Catalysts for the Oxygen Reduction Reaction. *Angewandte Chemie International Edition*. **9,** 60 (2021), pp. 4448-4463.

[12] G. Chen, P. Liu and Z. Liao. Zinc-Mediated Template Synthesis of Fe-N-C Electrocatalysts with Densely Accessible Fe-N$_x$ Active Sites for Efficient Oxygen Reduction. *Advanced Materials*. **8**, 32 (2020), pp. 1907399.

[13] C. H. Choi. The Achilles' heel of iron-based catalysts during oxygen reduction in an acidic medium. *Environmental Science*. (2018).

[14] E. Luo, Y. Chu and Liu. Pyrolyzed M-N$_x$ catalysts for oxygen reduction reaction: progress and prospects. *Environmental Science*. (2021).

[15] Niu, J. Wu, H. Xu and D. Cao. Hydrogen Peroxide Spillover on Platinum–Iron. *Journal of Catalysis*. 436 (2024).

[16] Z. Fan, H. Wan and H. Yu. Rational design of Fe-M-N-C based dual-atom catalysts for oxygen reduction electrocatalysis. *Chinese Journal of Catalysis*. 54 (2023), pp. 56-87.

[17] Y. Zhu, J. Sokolowski and X. Song. Engineering Local Coordination Environments of Atomically Dispersed and Heteroatom-Coordinated Single Metal Site Electrocatalysts for Clean Energy-Conversion. *Advanced Energy Materials*. **11**, 10 (2020).

[18] Y. X. Zhang, S. Zhang and Huang. General Synthesis of a Diatomic Catalyst Library via a Macrocyclic Precursor-Mediated Approach. *Journal of the American Chemical Society*. **8**, 145 (2023), pp. 4819-4827.

[19] X. Zhao, S. Sun and Yang. Atomic-Scale Evidence of Catalyst Evolution for the Structure-Controlled Growth of Single-Walled Carbon Nanotubes. *Accounts of Chemical Research*. **23**, 55 (2022), pp. 3334-3344.

[20] B. Liang, M. Su and Z. Zhao. Iron-involved zeolitic imidazolate framework-67 derived Co/Fe-NC as enhanced ORR catalyst in air–cathode microbial fuel cell. J*ournal of Electroanalytical Chemistry*. 962 (2024), pp. 118260.

[21] L. Li, W. Xie and J. Chen. ZIF-67 derived P/Ni/Co/NC nanoparticles as highly efficient electrocatalyst for oxygen reduction reaction (ORR). *Journal of Solid State Chemistry*. 264 (2018), pp. 1-5.

[22] H. Gao. Dual zeolitic imidazolate frameworks derived cobalt- and nitrogen-doped carbon nanotube-grafted flower- and leaf-like hierarchical porous carbon electrocatalysts for oxygen reduction. *Applied Surface Science: A Journal Devoted to the Properties of Interfaces in Relation to the Synthesis and Behaviour of Materials*. (2022).

[23] J. Gautam, T. D. Thanh and K. Maiti. Highly efficient electrocatalyst of N-doped graphene-encapsulated cobalt-iron carbides towards oxygen reduction reaction. *Carbon*. 137 (2018), pp. 358-367.

[24] X. X. Wang, D. A. Cullen and Y. Pan. Nitrogen-Coordinated Single Cobalt Atom Catalysts for Oxygen Reduction in Proton Exchange Membrane Fuel Cells. *Advanced Materials*. **11**, 30 (2018),

pp. 1706758.

[25] H. Huang, C. Zhang and H. Li. Carbon-coated FeCoNi nanocatalysts for the electrocatalytic oxygen reduction reaction. *Catalysis Science & Technology*. **9,** 14 (2024), pp. 2638-2645.

[26] H. Chang, J. Kang and L. Chen. Low-temperature solution-processable Ni(OH)$_2$ ultrathin nano sheet/N-graphene nanohybrids for high-performance supercapacitor electrodes. *Nanoscale*. **11**, 6 (2014), pp. 5960-5966.

[27] G. Zhang, F. Yin and G. Kofie. Preparation of N, S doped Co-Fe/C catalyst derived from ZIF-67/ionic liquid and its catalytic activity for oxygen reduction reaction in alkaline electrolyte. *Catalysis Today*. 441 (2024), pp. 114853.